# Dynamic Impedance of Two-Dimensional Superconducting Films Near the Superconducting Transition

Stefan J. Turneaure and Thomas R. Lemberger
*Dept. of Physics*
*Ohio State University*
*Columbus, OH 43210-1106*

John M. Graybeal[†]
*Dept. of Physics*
*University of Florida*
*Gainesville, FL 32611-8440*

The sheet impedances, $Z(\omega,T)$, of several superconducting $a$-$Mo_{77}Ge_{23}$ films and one $In/InO_x$ film have been measured in zero field using a two-coil mutual inductance technique at frequencies from 100 Hz to 100 kHz. $Z(\omega,T)$ is found to have three contributions: the inductive superfluid, renormalized by nonvortex phase fluctuations; conventional vortex-antivortex pairs, whose contribution turns on very rapidly just below the usual Kosterlitz-Thouless-Berezinskii unbinding temperature; and an anomalous contribution. The latter is predominantly resistive, persists well below the KTB temperature, and is weakly dependent on frequency down to remarkably low frequencies, at least 100 Hz. It increases with T as $e^{-U'(T)/kT}$, where the activation energy, $U'(T)$, is about half the energy to create a vortex-antivortex pair, indicating that the frequency dependence is that of individual excitations, rather than critical behavior.

[†] Present Address: Bell Laboratories, Lucent Technologies, 600-700 Mountain Avenue, Murray Hill, NJ 07974.

I. Introduction

For the past twenty years the superconducting to normal (S-N) transition in two-dimensional (2-D) films and Josephson junction arrays has been a very active area of research, revitalized by the quasi-two-dimensional nature of high-$T_C$ cuprate superconductors. The usual paradigm for identifying and discussing the transition is the static Kosterlitz-Thouless-Berezinskii (KTB) theory [1,2] and its extensions to dynamics [3,4,5]. These theories identify thermally excited vortex-antivortex pairs as the agents of dissipation, and focus on them, setting aside non-vortex (longitudinal) phase fluctuations and fluctuations in the amplitude of the order parameter. In this paper, we present a comprehensive study of the sheet impedance of superconducting films in zero field. Consistent with KTB, we find an abrupt drop in superfluid density and a concurrent rapid increase in sheet resistance at the KTB transition temperature. But below this temperature there is an anomalous impedance, primarily resistive, that is not present in KTB theory. This impedance becomes apparent at low frequencies (below 100 kHz) where the inductive impedance of the superfluid is small. Its frequency dependence is weak, but extends to surprisingly low frequencies, down to at least 100 Hz. It is possible that the S-N transition actually occurs well below the KTB unbinding temperature, and is mediated by a more subtle, longer-lived, excitation than the classic vortex-antivortex pair.

Many previous studies of the SN transition in 2-D were scaling analyses of nonlinear *dc* current-voltage (I-V) characteristics,[6-10] and thus cannot be compared directly with ours. While these papers generally acknowledge good agreement with KTB theory, Pierson et al.[11] have revisited the scaling analyses of many I-V measurements, as well as dynamic measurements on 2-D He films, and they find that when the dynamical exponent, z, is taken to be an adjustable parameter, the best scaling occurs for z = 5.6 ± 0.3, not the KTB value, z = 2. Pierson et al. discount the notion that deviations of I-V curves from KTB at low current densities arise from finite-size effects, or vortices generated by the ambient field, or by vortex pinning, and conclude that the 2-D S-N transition occurs below the universal KTB prediction, $T_C$. (A recent experimental and theoretical study of finite size effects in Josephson junction arrays can be found in ref. 10.)

There have been several previous studies of dynamics of 2-D superconductors, including indium-oxide [12] and *a*-MoGe films [13], wire networks [14], and Josephson junction arrays [15, 16]. Where our data overlap these studies, there is good agreement. In particular, all studies find that at temperatures below the KTB unbinding temperature, the sheet impedance is frequency dependent at remarkably low frequencies. The present work expands on these studies.

The frequency dependence of the sheet impedance provided by vortex-antivortex pairs has been explored theoretically in some detail. Ambegaokar and coworkers extended the static KTB theory in the context of vortex-antivortex dynamics in superfluid He films.[3] Halperin and Nelson extended this work to vortex-antivortex pairs in superconducting films.[4] Minnhagen [5] pointed out that Ambegaokar's expressions for the sheet conductivity violate the Kramers-Kronig relationship, and he has developed a model similar to that of Ambegaokar that remedies this problem. The Minnhagen phenomenology (MP) provides an expression for the vortex dielectric function, which determines the imaginary conductivity contributed by vortex-antivortex pairs. The real conductivity is determined by Kramers-Kronig transform.

The present work explores in detail the frequency and temperature dependence of the sheet impedance, $Z(\omega,T)$, of a model 2-D superconductor, amorphous MoGe, at normal-state sheet resistances $R_N$ up to 900 $\Omega$ and, for comparison, an amorphous-composite In/InO$_x$ film with $R_N$ near 4000 Ω. That these rather different materials share the same qualitative features indicates that the anomalous sheet impedance is general, not peculiar to a particular system. It follows that there exists an anomalous fluctuation that dominates the low-frequency behavior of Z below $T_C$, defined experimentally as the temperature where the sheet resistance and inductance increase very rapidly, consistent with vortex pair unbinding. This fluctuation must involve phase slips, hence vortices of some configuration. In Sec. IV we show that the conventional understanding of vortex-antivortex pairs does not capture their dynamics.

Before examining the data, it is useful to recall some of the principles and notations surrounding fluctuation effects. Thermal fluctuations become important in 2-D superconductors, (films and Josephson junction arrays), when $k_B T$ becomes comparable to the mean-field superconducting energy:

$$U_{00}(T) \equiv (\phi_0/2\pi)^2 \, L_0^{-1}(T), \tag{1}$$

where $\phi_0 \equiv h/2e$ is the flux quantum. The mean-field inverse sheet inductance, $L_0^{-1}(T)$, is proportional to the areal superfluid density, $n_{S0}(T)$, and it vanishes at the mean-field transition temperature, $T_{C0}$. $U_{00}(0)$ is typically an order of magnitude larger than $k_B T_{C0}$, so thermal fluctuations become important when $n_{S0}(T)$ is about an order of magnitude smaller than $n_{S0}(0)$, i.e., near $T = T_{C0}$. In 2-D films, with thickness $d << \xi$, where $\xi(T)$ is the G-L coherence length, $U_{00}(T)$ is the mean-field condensation energy, $V_C$ ($B_C^2/2\mu_0$), in a coherence volume, $V_C \equiv 4\xi^2(T)d$. One can also write: $U_{00} = \hbar^2 n_S(T)/4m$, where m is the electron mass, to make explicit the relationship between $U_{00}$ and superfluid density. In Josephson junction (JJ) arrays, $U_{00}$ is proportional to the mean-field Josephson coupling energy of one junction, the constant of proportionality being unity for square arrays and near unity for triangular and honeycomb arrays. When $U_{00}$ is written in terms of sheet inductance, as in Eq. (1), the expression is the same for films and arrays.

For our films, $U_{00}$ can be obtained by extrapolation of low temperature data using the weak coupling BCS result for $n_S(T)/n_S(0)$. A more useful energy, $U_0$, can be obtained from Eq. (1) by using the inductance of the background superfluid in place of the mean-field inductance. Since the contribution of vortex-antivortex pairs is small except very close to the unbinding temperature, below the critical region it is sufficient to calculate $U_0$ using the measured sheet inductance, exclusive of the anomalous part. In the critical region, $U_0$ is estimated to be about 40% larger than $U_{00}$.

Regardless of details, one would expect the S-N transition to occur near $k_B T/U_{00}(T) = 1$. The full KTB renormalization group theory predicts the S-N transition at $k_B T/U(T) = \pi/2$, where the superconducting energy, $U(T)$, is calculated *via* Eq. (1) but with the effective sheet inductance, $L(T)$, including vortex-antivortex pairs, in place of the mean-field inductance. Numerical simulations of square, triangular, and honeycomb JJ arrays find that $L_0/L$ is about 0.6 at the KTB transition temperature, so that the transition occurs at $k_B T_C/U_{00}(T_C) \approx 0.9$.[17] Consistent with this, we have found that the inverse sheet inductance of the *a*-MoGe films discussed below drops precipitously at $k_B T/U_{00}(T) \approx 0.9$, when the anomalous component of the sheet impedance is set aside.[18] And fluctuations suppress the inverse sheet inductance to about 60% of its mean-field value just before the rapid drop. On this basis, we argue that the impedance of our films should be interpreted as the expected impedance of superfluid plus vortex pairs, plus an anomalous contribution.

It is straightforward to identify the anomalous impedance. Below $T_C$, as discussed below, vortex-antivortex pairs should be inductive at our experimental frequencies, so all of the sheet resistance, $R(\omega,T)$, is anomalous. The anomalous part, $L_a$, of the experimental sheet inductance, $L(\omega,T)$, can be identified from its dependence on frequency, since the inductance of the background superfluid plus bound vortex-antivortex pairs, $L_{SF}(T)$, is independent of frequency at temperatures and frequencies of interest here. Theories focus on the sheet conductance, $G(\omega,T) = \sigma_1(\omega,T)d - i\sigma_2(\omega,T)d \equiv Z^{-1}(\omega,T)$, especially the peak in $G_1(\omega,T)$ vs. T, so we present our data in this form, too. But in our view, at temperatures below the very narrow critical region, the impedance, $Z(\omega,T)$ is more transparent because the impedances of the superfluid background and thermal vortex pair excitations are in series, in analogy with the impedance of pinned vortices.[19]

The outline of this paper is as follows. In Sec. II, the experimental method and sample properties are discussed. In Sec. III, experimental results are presented and features common to all films are highlighted. Section IV outlines conventional properties of individual vortex-antivortex pairs. Sec. V then argues that unbinding of conventional pairs is responsible for the rapid changes in Z at and above the experimental $T_C$, but cannot account for the anomalous impedance below $T_C$.

II. Experimental Method and Sample Properties.

The sheet impedance, $Z(\omega,T)$, was determined at frequencies, $f = \omega/2\pi$, from 100 Hz to 100 kHz using a two-coil mutual inductance technique with the drive and pick-up coils coaxial and located on opposite sides of the film.[20] By means of a look-up table containing over 10,000 pairs of M and Z values, calculated for the geometry of the actual film and coils, the real and imaginary parts of Z were determined from the in-phase and quadrature components of M. Great care was taken to ensure that Z was independent of the excitation amplitude. Measurements also were performed as a function of perpendicular magnetic field to identify the field range over which Z was independent of field.[21] All data presented here were taken with the field was nulled sufficiently so as not to affect the film's impedance.

Table I lists film properties. $a$-Mo$_{77}$Ge$_{23}$ films with thicknesses down to 21.5 Å were grown on oxidized Si, with 100 Å Ge buffer layers below and above. The beauty of these films is that, fluctuations aside, they are nearly perfect weak coupling BCS superconductors in the respect that, as shown in refs. 19 and 21, their mean-field sheet inductances have the same BCS T-dependence, that is, $L^{-1}_{MF}(T/T_{C0})/L^{-1}_{MF}(0)$ vs. $T/T_{C0}$ is the same for all, even though $L^{-1}_{MF}(0)$ and $T_{C0}$ vary substantially with thickness. Data also are presented for a 10 mm diameter In/InO$_x$ film (Film $I$ in Table I). The superfluid density for this film does not follow the weak coupling BCS theory as well as MoGe films do, and in that sense it is less ideal.

III. Experimental Results

In this section data are presented for the complex conductivity and impedance of several films. We emphasize that a great deal of effort went into ascertaining that data were taken in the linear response regime, where the sheet impedance was independent of the magnetic field produced by current in the drive coil, and into checking that the residual ambient field was negligibly small.[21]

A few comments about experimental uncertainties are in order. The experiment measures the magnetic field produced by induced currents in the sample. The current density does not vary through the thickness of our very thin films. Assuming that the film is homogeneous on a length scale much shorter than the 1 mm radii of the drive and pickup coils, the experiment yields $Z(\omega,T)$ directly. Uncertainty in the film thickness, d, enters only when we calculate the resistivity, $\rho \equiv Zd$. Signal-to-noise decreases as $\omega$ decreases because the measured pick-up voltage is proportional to $\omega$. Signal-to-noise decreases as T exceeds $T_C$ and the field produced by currents in the sample becomes small. Uncertainty in R grows near the low temperature tail in $\sigma_1$. Here small uncertainties (less than $1°$) in the phase of the mutual inductance are responsible. Since the dissipation peak extends to lower temperatures as $\omega$ is reduced, the sheet resistance is known with less accuracy for high frequencies at lower temperatures.

Figure 1 shows $\mu_0\omega\sigma$ vs. T and Fig. 2 shows $L^{-1}$ and R at 190 Hz $\leq$ f $\leq$ 50 kHz for *a*-MoGe Film C. Consider data at 50 kHz. At about 4.92 K, R begins to increase very rapidly, and $L^{-1}$ begins to drop. We define this temperature to be the experimental $T_C$, and we associate it with the unbinding of vortex-antivortex pairs. Below $T_C$, R is small, less than $10^{-7}$ $R_N$, but it should be immeasurably small. Figure 3 shows that for $T < T_C$, R exhibits activated behavior with an activation energy of $3.5U_0(T)$, about half the energy needed to create a vortex-antivortex pair, as discussed below. It is possible that R vanishes at a phase transition well below $T_C$, but if so, the transition occurs when R is below our sensitivity. Finally, we note that R depends weakly on frequency down to 190 Hz, with no sign of saturation.

$L^{-1}$ increases slowly with $\omega$. $L_a$ can be extracted by fitting $\omega L$ with an ordinary inductive term, $\omega L_{SF}(T)$, which is strictly proportional to $\omega$ and includes the inductances of the superfluid and conventional vortex pairs, and an anomalous part, $\omega L_a(\omega,T)$, with a power law frequency dependence. It turns out that R and $\omega L_a(\omega,T)$ have the same power law dependence over several decades of frequency, and therefore are consistent with being Kramers-Kronig transforms of each other. This lends confidence to the separation of L into two components. Results for MoGe films F and G are similar to C.[21]

Similar results are found for In/InO$_x$ Film I. Figure 4 shows $\mu_0\omega\sigma_1$ and $\mu_0\omega\sigma_2$ at 50 kHz and the normal-state resistance, $R_N(T)$. $R_N \rightarrow 0$ at the same temperature that superfluid appears, indicative of an electrically homogeneous film, even though the microstructure is an amorphous composite. Figure 5 shows $\mu_0\omega\sigma(\omega,T)$ and Fig. 6 shows $R(\omega,T)$ and $L^{-1}(\omega,T)$ for 200 Hz $\leq$ f $\leq$ 100 kHz. The important qualitative features are the same as for MoGe films. At 50 kHz, R begins to increase rapidly at the same temperature ($T_C \approx 2.685$ K) where $L^{-1}$ begins to drop, and there is an anomalous impedance below $T_C$ that is frequency dependent down to at least 200 Hz. We note that ref. 12 found a similar frequency dependence for $L^{-1}$ in an In/InO$_x$ film. They did not present data for R. A minor quantitative difference between In/InO$_x$ and a-MoGe is that for In/InO$_x$, $R(\omega,T)$ increases with T with an activation energy of about 2.2 $U_0(T)$.

The frequency dependence of the anomalous impedance is weak and extends to remarkably low frequencies. To explore frequency dependence in detail, measurements were made at fixed T for MoGe Film C while sweeping the frequency. Noise at low $\omega$ was reduced by averaging thousands of measurements over periods of about 10 minutes. Figure 6 shows results at four temperatures. The top panel shows that L decreases and approaches a constant as $\omega$ increases. The inset of the top panel shows a fit of $\log(L - L_{SF})$ to: const. + (b-1) $\log(f)$, where $L_{SF}$ was adjusted to obtain the best straight line. The best fit has b = 0.125, so the anomalous reactance is: $\omega L_a = A(T)\omega^{0.125}$. The bottom panel in Fig. 6 shows R vs. f (solid symbols) at four temperatures, and $\omega L_a$ vs. f at T = 5.406 K (open circles). For all temperatures, R and $\omega L_a$ are proportional to $\omega^b$, with b = 0.13 $\pm$ 0.02. For In/InO$_x$ Film I, b $\approx$ 0.20$\pm$0.05. Our main point is that $Z_a$ is a weak function of $\omega$ at surprisingly low frequencies.

Figure 7 illustrates that $L_{SF}^{-1}$ is not much larger than $L^{-1}$ measured at the highest experimental frequency, typically 50 or 100 kHz. The lower curves are $L^{-1}(T)$ measured at 10 kHz and 50 kHz, and the dashed curve is $L_{SF}^{-1}(T)$. For $T < T_C$, the dashed curve is quite close to the 50 kHz data, so errors in extrapolation are small.

## IV. Expected Behavior of Vortex-Antivortex Pairs.

In this section, we construct a simple description of vortex-antivortex pairs that provides estimates for various important parameters such as the density of vortices at the unbinding transition and the width of the critical region. It supports our conclusions that the unbinding transition for conventional vortex-antivortex pairs occurs as expected, and that there is an anomalous dissipative mechanism that cannot be described with conventional vortex-antivortex pairs. Since interactions among pairs are important only in a very narrow region near the unbinding transition,[22] and our main focus lies below this region, we will include interpair interactions in the simplest fashion. Our estimations for vortex-pair properties are more accurate if we calculate the characteristic superconducting energy introduced in Eq. (1) by replacing the mean-field sheet inductance with the inductance of the background superfluid, which is suppressed by nonvortex thermal phase fluctuations, but is smooth through the pair-unbinding transition. We denote this energy as $U_0$.

To begin, we review how fluctuation effects evolve with increasing temperature. At a low temperature, say, $k_B T/U_0(T) \approx 1/20$, small amplitude, nonvortex phase fluctuations suppress the background superfluid density, $n_{S,B}(T)$, below the mean-field density, $n_{S0}(T)$, by perhaps a percent,[23] and the areal density, $n_p(T)$, of vortex pairs is negligible. As T increases, nonvortex phase fluctuations increase in intensity, ultimately suppressing $n_{S,B}$ to 70 - 75% of $n_{S0}$ at the unbinding transition. These background phase fluctuations are an essential part of the story because they are responsible both for generating vortex-antivortex pairs and for driving their Brownian motion. They are considered in some detail in ref. (24). For reference, assuming $n_{S,B}(T) \approx 0.75\ n_{S0}(T)$ at the unbinding transition, KTB theory predicts a transition at $k_B T/U_0(T) \approx .75\ \pi/2 \approx 1.2$.

To estimate the density of vortex pairs, we need their energy. The calculation is straightforward. The energy has three portions: (1) $E_{M,pair}$, associated with suppression of, and gradients in, the Magnitude of the order parameter; (2) $E_{K,pair}$, the kinetic energy of supercurrents, and (3) magnetic field energy, which is negligible. $E_M$ is given by [25,26]:

$$E_M = (\pi U_0/2) \int_0^\infty dr\ r\ [(1-f^2)^2 + 2(\nabla f)^2], \qquad (2)$$

where $f(r)$ is the normalized order parameter. For a single vortex, $E_M = 2.45 U_0$. [Ref. 27 used half of this value.] The energy $E_K$ is calculated from the sheet supercurrent density, $K_S(r)$:

$$E_K = \pi \mu_0 \lambda_\perp \int_0^\infty dr\ r\ f^2(r)\ K_S^2(r), \qquad (3)$$

where $\lambda_\perp \equiv \lambda^2/d$. Within the London model for a single vortex, the sheet supercurrent density $K_S(r)$ outside the core is proportional to $1/r$ for $\xi << r << \lambda_\perp$, and proportional to $1/r^2$ for $r >> \lambda_\perp$.[28] To get $K_S(r)$ and $f(r)$ in the core, we solved the G-L differential equation numerically as has been done previously [25].

To calculate the energy of a vortex and antivortex separated by a distance $\rho$, one should solve the G-L differential equations, but this has not been accomplished since the axial symmetry of one vortex is broken for a pair. $E_{M,pair}$ is approximately twice $E_M$ for a single vortex. To determine $E_{K,pair}$, we integrated the kinetic energy density of the supercurrents associated with a pair,[21] assuming that the supercurrent patterns for a single vortex could be added. Figure 8 shows the calculated energy of a pair as a function of $\rho$. To a good approximation, $E_{pair}$ increases logarithmically with $\rho$:[4,5]

$$E_{pair} = E_{C,pair}(r_0) + 2\pi U_0 \ln(\rho/r_0), \tag{4}$$

where $r_0$ is the size of the smallest pair that is well defined. We take $r_0 = 2\xi(T)$. Figure 8 shows that Eq. (4) is a very good approximation for $\rho > 1.5r_0$. Pairs smaller than $2\xi$ are effectively fluctuations in the order parameter amplitude, since there is very little current associated with them. Given that the energy of minimum sized pairs is uncertain we take $E_{C,pair}(r_0)$ to be the value of the logarithmic asymptote (dashed line in Figure 8) evaluated at $r_0$. For $r_0 = 2\xi$, $E_{C,pair} = 6.22U_0$. This procedure gives the correct energy for pairs larger than about $2r_0$ and much smaller than $\lambda_\perp$, which is the range of interest here.

The density of vortex pairs, $n_p$, can be estimated by assuming that all pairs are of the minimum size, $r_0$, and calculating the probability of finding a pair in each $2r_0 \times 2r_0$ cell of the film:

$$n_p = [1/4r_0^2] \, N_0 \, e^{-E_{C,pair}(r0)/k_B T} / [1 + N_0 \, e^{-E_{C,pair}(r0)/k_B T}], \tag{5}$$

where $N_0$ is the number of independent ways that the pair can be oriented in the cell. Roughly, the vortex can be in any quadrant of the cell, and the antivortex can be in any of the other three quadrants, so $N_0 \approx 12$. Certainly, $N_0$ should be much larger than unity. Our conclusions are insensitive to its precise value. The fraction, $f_N$, of the film area that is "normal" is approximately the fraction occupied by vortex cores: $f_N \approx n_{p0} \, 4\xi^2$. We would expect $f_N$ to be roughly one percent at the transition; certainly it must be much less than the 2-D percolation value of 50%, and to decrease very rapidly below the transition. Consistent with this expectation, from Eq. (5) we estimate $f_N = 0.001$ at $kT/U_0 \approx 0.7$, and $f_N = 0.01$ at $kT/U_0 \approx 1$ with parameters, $r_0 = 2\xi$, $E_{C,pair} = 6U_0$ and $N_0 = 12$. Thus, just above the unbinding transition, where all pairs are unbound and resistive, we expect the sheet resistance to be about 5% of $R_N$, a reasonable value. We will use Eq. (5) with the estimated impedance of vortex pairs to compare to the measured sheet impedance.

We now show that two simple estimates of the pair unbinding transition temperature are very close to the measured $T_C$. An upper limit is where the *rms* size of noninteracting pairs diverges. A better, lower, upper limit comes from the temperature where noninteracting pairs overlap. KTB theory includes interpair interactions which facilitate pair unbinding, and therefore obtains a still lower transition temperature.

The probability that a given pair has a separation $\rho > r_0$ is determined by the increase in free energy with separation: $(2\pi U_0 - k_B T)\ell n(\rho / r_0)$, which leads to an *rms* pair size: [1]

$$\langle \rho^2 \rangle^{1/2} / r_0 = [\pi U_0/k_B T - 1]^{1/2} / [\pi U_0/k_B T - 2]^{1/2}, \tag{6}$$

that reaches 2 at $k_BT/U_0(T) \approx 1.4$, and diverges at $k_BT/U_0(T) = \pi/2$. At a slightly lower temperature, pairs overlap so much that it is impossible to say which vortex is paired with which antivortex, so vortices are effectively unbound. If we take the criterion to be $[n_p\langle\rho^2\rangle]^{1/2} \approx 0.4$, then with $N_0 = 12$ and $E_{C,pair} = 6U_0$, unbinding occurs at $k_BT/U_0(T) \approx 1.3$. KTB theory finds an unbinding transition at $k_BT/U_0(T) \approx 1.2$, while the better of our two upper-limit estimates puts the transition at $k_BT/U_0(T) \approx 1.3$. These are quite close, and we conclude that pairs unbind at a temperature very close to that predicted by KTB, even if details of the KTB theory were incomplete. The critical region occupies temperatures, $1.2 < k_BT/U_0(T) < \pi/2$. At higher temperatures, all pairs are unbound. If $U_0 \approx 0.7\, U_{00}$, then for our films the critical region extends approximately one-third of the way from the unbinding transition at $T_C$ to the mean-field transition at $T_{C0}$.

A typical experimental resistance at the upper edge of the critical region is perhaps a few tenths of a percent of the normal-state resistance.[7] This suggests that at the transition, vortex cores occupy somewhat less than 1% of the film area. $N_0$ in Eq. (5) should be closer to 3 than to 12.

In JJ arrays vortex-pair and nonvortex fluctuations together suppress the inverse inductance to about 60% of its mean field value just below the unbinding transition. If nonvortex fluctuations suppress the inverse inductance of the background superfluid to 75% of its mean-field value, then the inductance of vortex-antivortex pairs is about 25% of the measured inductance. Our simple model is consistent with this result. First, we note that a typical vortex-antivortex pair is small. The probability, $P(r)$, that a given pair has a separation greater than r is:

$$P(r) = (r_0/r)^{2\pi U_0/kT - 2}. \qquad (7)$$

Thus, at $k_BT/U_0 = 1$, the probability for a pair to be ten times larger than its minimum size, i.e., $\langle\rho^2\rangle^{1/2} / r_0 = 10$, is about $10^{-4}$. The number of unbound pairs, with $\langle\rho^2\rangle^{1/2} \approx \lambda_\perp \approx 1000\, r_0$, is negligible for $k_BT/ U_0 < 1$.

We estimate the impedance of a bound pair as follows. A pair with separation, $\rho$, has a dipole moment, $\rho\phi_0$, and polarizes in response to the average background supercurrent, $\mathbf{K}_s$, (taken to be along the x direction and sinusoidal in time). The impedances of the superfluid and vortex pairs are in series as long as the pairs are not too close together, which is the simplified case under consideration here. The *ac* supercurrent requires an average electric field, $E_{s,x} = -i\omega L_s K_{s,x}$, where $L_s$ is the inductance of the superfluid. In thermal equilibrium, each dipole has an average y component: $[(\rho\phi_0)^2/k_BT]\, K_s$. The net polarization is proportional to $K_{s,x}$, hence to $E_{s,x}(\omega)/i\omega$:

$$P_y(\omega) = \chi_p(\omega)\, E_{s,x}(\omega) / -i\omega, \qquad (8)$$

where the low-frequency susceptibility is:

$$\chi_p = n_p \langle\rho^2\rangle\phi_0^2 / [L_s k_BT(1 - i\omega\tau)]. \qquad (9)$$

$\tau$ is an average equilibration time that depends on $\langle\rho^2\rangle^{1/2}$. The average electric field due to polarizing pairs arises from their velocity along y:

$$E_{p,x} = -i\omega P_y(\omega) = \chi_p E_{s,x}, \tag{10}$$

so the sheet impedance is:

$$Z(\omega,T) = E_{s,x} (1 + \chi_p) / K_{s,x} = -i\omega L_s \varepsilon_p, \tag{11}$$

where the real and imaginary parts of the inverse vortex pair dielectric function, $\varepsilon_p^{-1}$, are related by Kramers-Kronig transform.[5]  The pair impedance,

$$Z_p = R_p(\omega,T) + i\omega L_p(T) \equiv E_{p,x} / K_{s,x} \approx i\omega n_p \langle\rho^2\rangle \phi_0^2 / k_B T(1 - i\omega\tau), \tag{12}$$

is inductive at $\omega \ll 1/\tau$ and resistive for $\omega \gg 1/\tau$.

$\tau$ is the time for a typical vortex-antivortex pair to sample all possible orientations relative to $K_S$, i.e., $\tau \approx D/\langle\rho^2\rangle$, where D is the vortex diffusion constant. With the expression, $D = 28e^2\xi^2 k_B T R_N / \hbar^2\pi = (7\xi^2/\pi)(kT/\hbar)(R_N/R_Q)$ [3,4], with $R_Q \equiv \hbar/4e^2 \approx 1$ k$\Omega$, and $\langle\rho^2\rangle^{1/2} \approx 2\xi$, we have:

$$1/\tau \approx (7/\pi)(k_B T/\hbar)(R_N/R_Q). \tag{13}$$

With typical values, $R_N = 300$ $\Omega$ and T = 5 K, we find $1/\tau \approx 10^{11}$ rad/s, which is much larger than our maximum experimental $\omega$, $6\times10^5$ rad/s, so vortex pairs are inductive.

For film C, $k_B T/U_0(T) \approx 1.0$ at 4.903 K. $L_{SF}$ is about 1.2 nH so the contribution of vortex pairs is about 0.3 nH. From Eq. (12) and $n_p\langle\rho^2\rangle \approx 0.01$, the estimated inductance of vortex pairs is about 0.6 nH. Of course, $n_p\langle\rho^2\rangle$ may be an order of magnitude smaller than 0.01, and the estimated inductance of pairs would then be near 0.06 nH. Given the uncertainties, we consider this to be good agreement and further confirmation that the drop in $L_{SF}$ represents the unbinding of a low density of conventional vortex-antivortex pairs. Figure 10 shows that the drop in $L_{SF}^{-1}$ is consistent with KTB theory.

Finally, we show that the resistance of vortex pairs should be much smaller that what we observe. Motion of pairs subject to an *ac* supercurrent would cause dissipation due to viscosity, and give rise to a small resistance, $R_p$. From Eq. (13):

$$R_p/\omega L_p \approx 7\pi^3 (\hbar\omega/k_B T) R_Q / R_N. \tag{14}$$

Clearly the quadratic frequency dependence predicted for $R_p$ is stronger than is observed for R below the unbinding transition. For a quantitative comparison, we need to pick a particular frequency, arbitrarily taken to be 50 kHz. For film C, Eq. (14) yields $R_p /\omega L_p \approx 10^{-4}$. If we take $L_p = 0.3$ nH, then $R_p$ should be about $10^{-8}$ $\Omega$, which is four orders of magnitude smaller than the measured sheet resistance just below the transition. Thus, the dissipation of conventional pairs is undetectable at our measurement frequencies. The anomalous impedance below $T_C$ remains to be explained.

V. Discussion of $Z_a(\omega,T)$.

Experimentally, the anomalous sheet impedances, $Z_a(\omega,T) = R(\omega,T) + i\omega L_a(\omega,T)$ of *a*-MoGe and In/InO$_x$ films are quite similar in their most important features: 1) Over the experimental frequency range, 100 Hz to 100 kHz, $Z_a$ has a weak dependence on $\omega$, being roughly proportional to $\omega$ to a small T-independent power; 2) consistent with the weak frequency dependence, $Z_a$ is mostly resistive: $R/\omega L_a \approx 8$, independent of $\omega$ and T; 3) $Z_a$ has an Arrhenius T dependence, with an excitation energy of about $3.5U_0$ in *a*-MoGe and $2.2U_0$ for In/InO$_x$. These similarities argue that the observed behavior is generic to 2-D superconductors, and not due to microstructure. Another similarity whose significance is unclear is that $R(\omega,T) \approx$ 1 m$\Omega$ just below T$_C$, independent of normal state resistance.

Since the anomalous impedance is dissipative, it involves vortices and antivortices in some configuration. The Arrhenius T-dependence of $Z_a(\omega,T)$ suggests that the dissipative excitations are not interacting, so the frequency dependence is intrinsic to each excitation and not a result of critical behavior. The frequency dependence is the most puzzling feature because it persists below the conventional unbinding transition, and to such low frequencies. On the first point, we note that we were able to measure a nonzero sheet resistance down to $k_BT/U_0(T) \approx 1/3$, well below the conventional transition at $k_BT/U_0(T) \approx 1.2$. When Pierson et al.[11] reanalyzed the I-V data of van der Zant et al.[29] on a Josephson junction array, allowing the dynamical exponent z to be a free parameter, they found that the transition occurred at $k_BT/U_0(T) \approx 0.5$, also well below the conventional transition. Data at lower frequencies and temperatures are needed to see whether there is an S-N transition below the unbinding temperature.

On the second point, we note that the excitation energy of about $3.5U_0$ suggests that the anomalous excitation has spatial dimensions of a single vortex, a few coherence lengths. Characteristic times associated with this distance are much shorter than the experimental time scale of 10 ms. The time for an electron or phonon to travel a coherence length ballistically is very short. Vortex diffusion sets a time scale of $\hbar/k_BT$ for typical films, which is very short. Another characteristic vortex time is the time for a vortex-antivortex pair to annihilate, in the absence of Brownian perturbations. This time is proportional to the square of their initial separation, $\rho(0)$: $\tau_{annih} \approx (L_{SF}/R_N)[\rho(0)/\xi(T)]^2$, which is about equal to $\hbar/k_BT_C$ for our films, and is also too short to account for frequency dependence at 100 Hz.

Other experiments have observed similar anomalous low-frequency behavior in films: Fiory et al. [12] down to 14 Hz in an In/InO$_x$ film similar to ours, and Festin et al.[30] down to 0.1 Hz in a YBCO film. The phenomenon is not confined to continuous films. In a triangular Josephson junction array at $k_BT/U_0 \approx 1/2$, (T = 3.27 K; T$_C \approx$ 3.70 K; R$_N$/L $\approx 3\times10^6$ rad/s), Theron et al. [15] observed a 40% increase in sheet inductance (0.7 to 1 nH) as frequency decreased from 10 to 0.16 kHz. For comparison, in film C at $k_BT/U_0 \approx 1/2$, (T = 4.83 K; T$_C$ = 4.92 K; R$_N$/L $\approx 3\times10^{11}$ rad/s), we observed an increase of 70% (0.9 to 1.5 nH) as frequency decreased from 10 kHz to 0.19 kHz. The quantitative similarity is striking, considering the physical differences between films and arrays. Theron et al. concluded that the diffusion of field-induced vortices in their arrays was anomalously sluggish.

It has been suggested that the anomalous excitation might be single thermally excited vortices created at the edges of the films.[32] We do not have a model for the dynamics of these vortices, so further measurements would be needed to assess their contribution to the measured sheet impedance.

VI. Conclusion.

   The sheet impedances of 2-D superconducting $a$-MoGe and In/InO$_x$ films exhibit features expected from the presence of thermally excited vortex-antivortex pairs, especially, rapid increases in resistance and inductance at the same temperature due to pair unbinding. In addition, below the unbinding transition, there exists an anomalous, dissipative excitation with dynamics that extend to frequencies well below characteristic frequencies for vortex pairs. The anomalous excitation seems to exist in arrays of Josephson junctions, as well as films. Its Arrhenius T dependence suggests that the slow dynamics are a property of individual excitations rather than arising from interactions among conventional vortex-antivortex pairs. Identifying this excitation remains as an important challenge to the community. Understanding this excitation will improve our insight into the T=0 superconductor-to-insulator transition and bolster confidence in our ability to interpret the sheet impedance of cuprate superconductors.

*Acknowledgments:*  This work was supported in part by DoE grant DE-FG02-90ER45427 through the Midwest Superconductivity Consortium. JMG gratefully acknowledges research support from the NHMFL In-House Research Program, under NSF/DMR contract # 9527035 and SJT gratefully acknowledges support from an Ohio State University fellowship and technical assistance from Margarita Rokhlin and Dr. Saad Hebboul.

**TABLE CAPTION**

I. Film Parameters. Films C, F and G are amorphous $a$-$Mo_{77}Ge_{23}$, and Film I is amorphous-composite $In/InO_x$. d is the nominal film thickness. $L^{-1}(0)$ is the measured inverse sheet inductance extrapolated to T = 0. The normal state sheet resistance, $R_N(15 \text{ K})$, is nominal for the MoGe films [31] and measured for the $In/InOx$ film. The uncertainty in $T_{C0}$ is about 15% of $T_{C0}$ – $T_C$ for the MoGe films and somewhat larger for the $In/InO_x$ film.

Table I.

| Film | C | F | G | I $^*$ |
|---|---|---|---|---|
| d  (Å) | **46** | **27.5** | **21.5** | **190** |
| $L^{-1}(0)$  $(nH)^{-1}$  (±4%) | **9.55** | **4.21** | **2.57** | **0.692** |
| $R_N$  (Ω)  (±5%) | **387** | **674** | **885** | **4150** |
| $T_{C0}$ (K) | **5.043** | **3.881** | **3.167** | **3.048** |
| $T_C$ (K) (±5 mK) | **4.920** | **3.734** | **2.999** | **2.685** |
| $(T_{C0} – T_C)/T_{C0}$ (±15%) | **0.024** | **0.038** | **0.053** | **0.12** |

FIGURE CAPTIONS

1. T dependence of $\mu_0\omega\sigma$ for MoGe Film C measured at $f = 0.19, 1, 2, 5, 10$ and 50 kHz. $\mu_0\omega\sigma_2$ increases monotonically with frequency, and $\mu_0\omega\sigma_1$ peaks at higher temperatures as frequency is increased.

2. T dependencies of $L^{-1}$ and R for MoGe Film C, calculated from data in Fig. 1.

3. $\ln(R)$ vs. $U_0/k_BT$ for Film C at $f = 50$ kHz. The linear fit indicates that $R(T)$ is Arrhenius with an activation energy of $3.5U_0$.

4. $\mu_0\omega\sigma$ measured at 50 kHz and the normal state sheet resistance for In/InO$_x$ Film I. The uncertainty in both $R_N$ and $\mu_0\omega\sigma_2$ is about 10%.

5. T dependence of $\mu_0\omega\sigma$ for In/InO$_x$ Film I at $f = 0.2, 0.5, 1, 2, 5, 10$ and 50 and 100 kHz. $\mu_0\omega\sigma_2$ increases monotonically with frequency near the transition, and $\mu_0\omega\sigma_1$ peaks at higher temperatures as frequency is increased.

6. T dependencies of $L^{-1}$ and R for In/InO$_x$ Film I, calculated from data in Fig. 5.

7. Top panel shows L vs. f for four temperatures. The inset shows that $L_a = L - L_{SF} \propto \omega^{-0.875}$. The bottom panel shows R (filled symbols) for the same 4 temperatures as well as $\omega L_a$ for one temperature (open circles). All five of these impedances are proportional to $\omega^{0.13\pm0.02}$.

8. Example of the extrapolation procedure used to obtain $L^{-1}_{SF}(T)$ for Film C. The solid lines are measurements $f = 10$ kHz and 50 kHz. The dashed line is the extrapolation to high frequency, which yields $L^{-1}_{SF}$.

9. Calculated energy of a vortex-antivortex pair in a high $\kappa$, 2-D superconductor. $E_{M,pair}$ is the energy from suppression and gradients of the Magnitude of the order parameter, $E_{K,pair}$ is the Kinetic energy of the supercurrents associated with the pair. The dashed line is the logarithmic asymptote of the pair energy, and its value at $r = 2\xi$ yields a core energy of $6.22U_0$ for a pair of minimum size, $r = 2\xi$.

10. $L_{SF}(0) / L_{SF}(T/T_{C0})$ vs. $T/T_{C0}$ for four films. The intersection of dashed line and data is where the KTB vortex-pair unbinding transition is predicted to occur.

Figure 1.

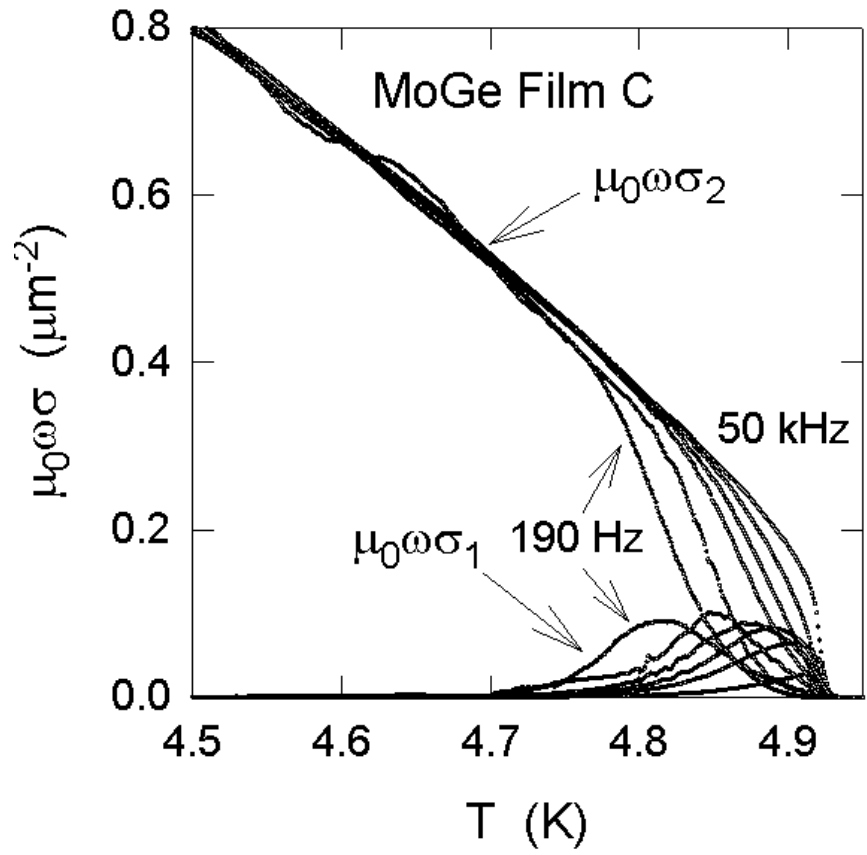

Figure 2.

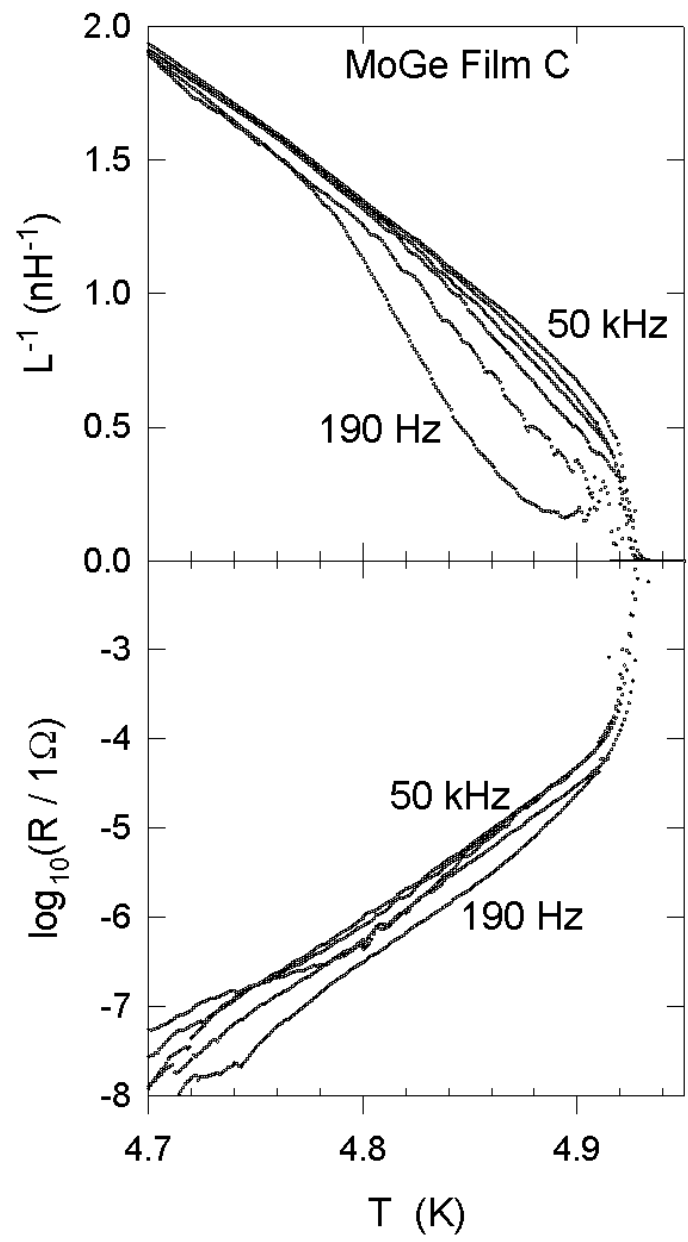

Figure 3.

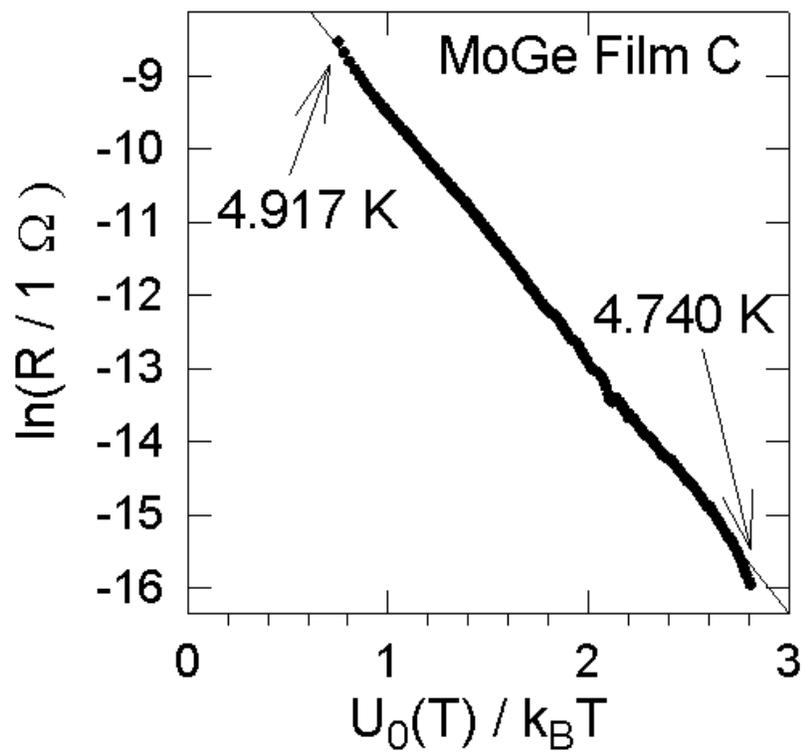

Figure 4.

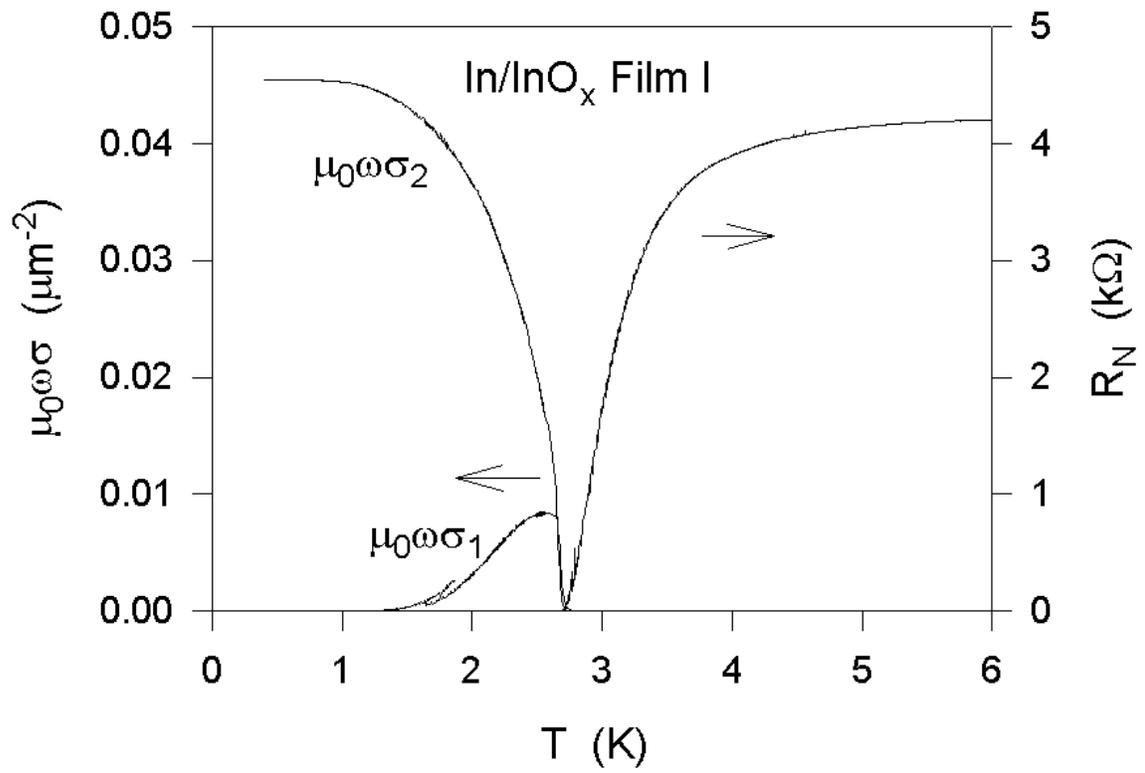

Figure 5.

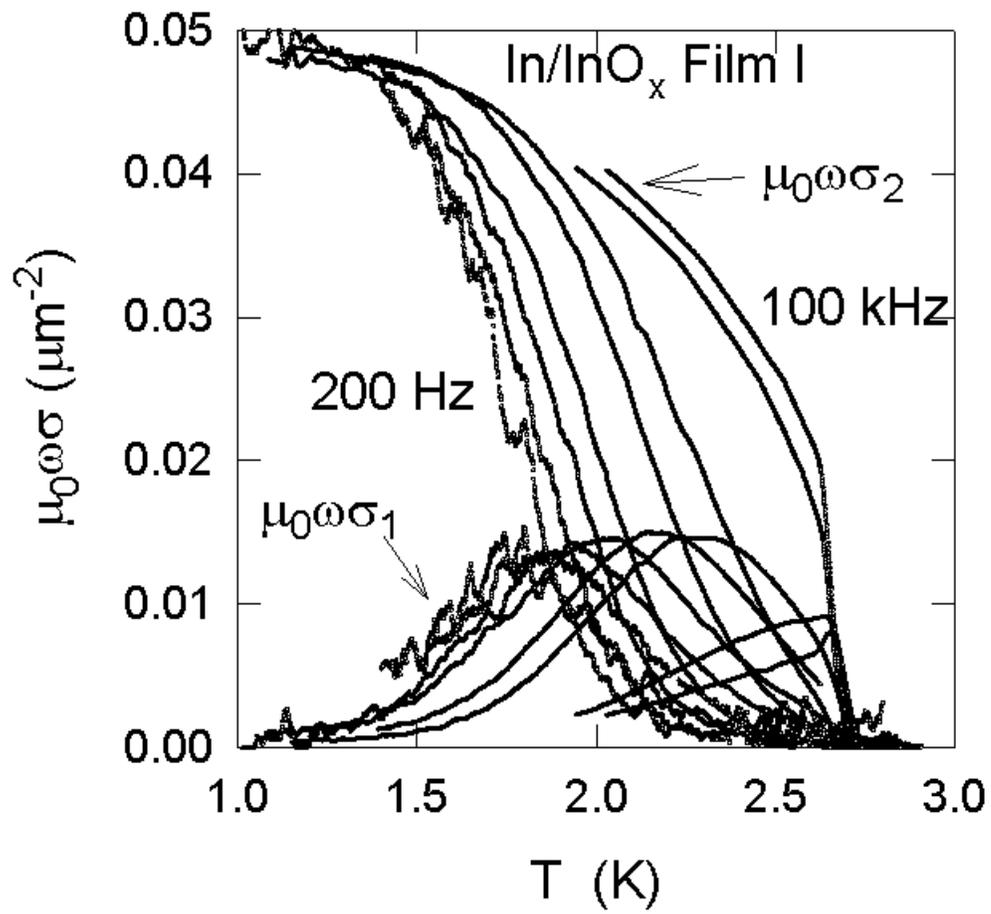

Figure 6.

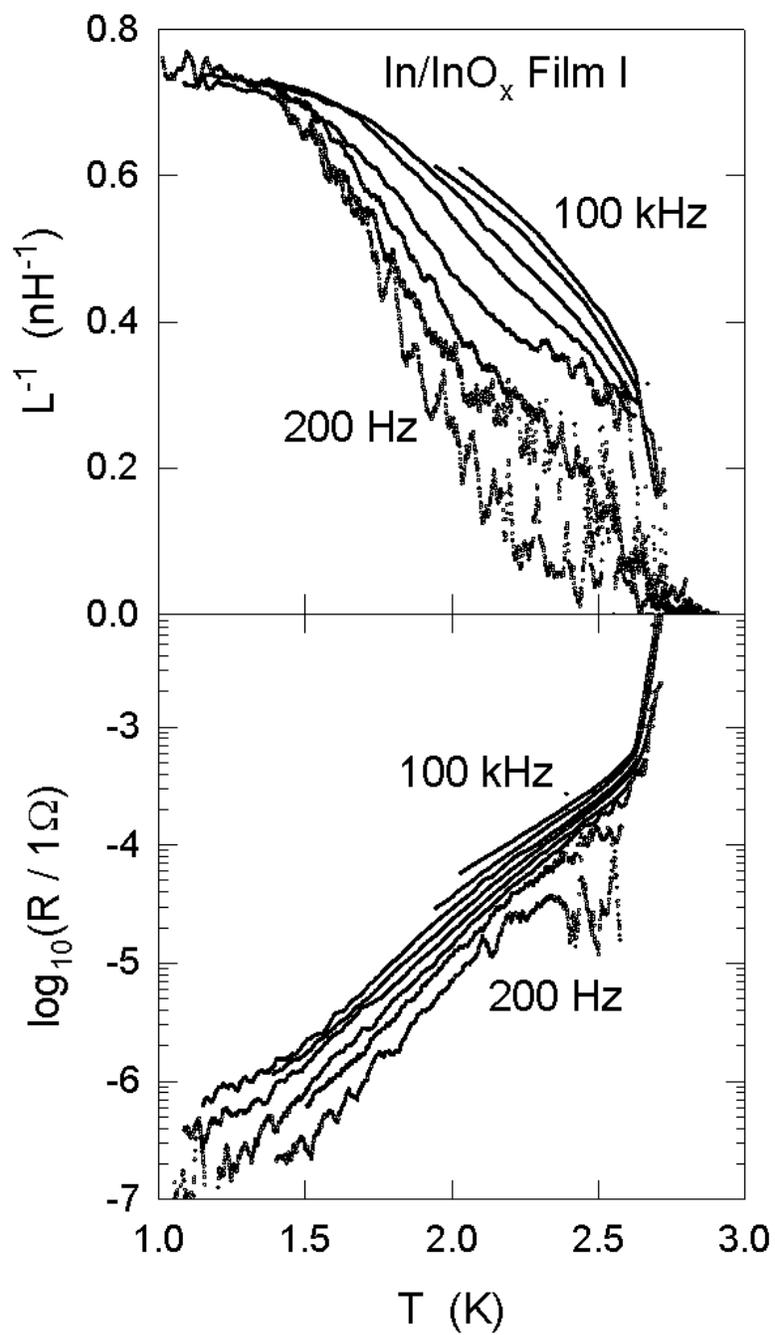

Figure 7.

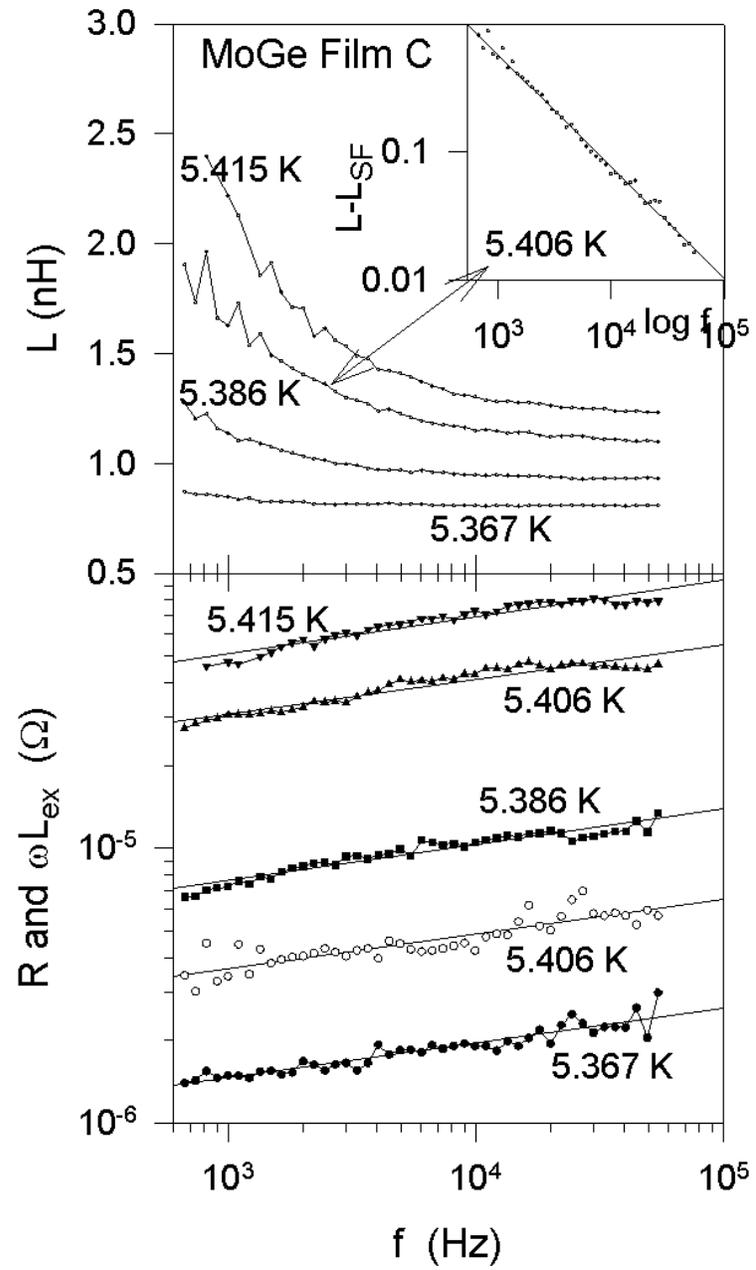

Figure 8.

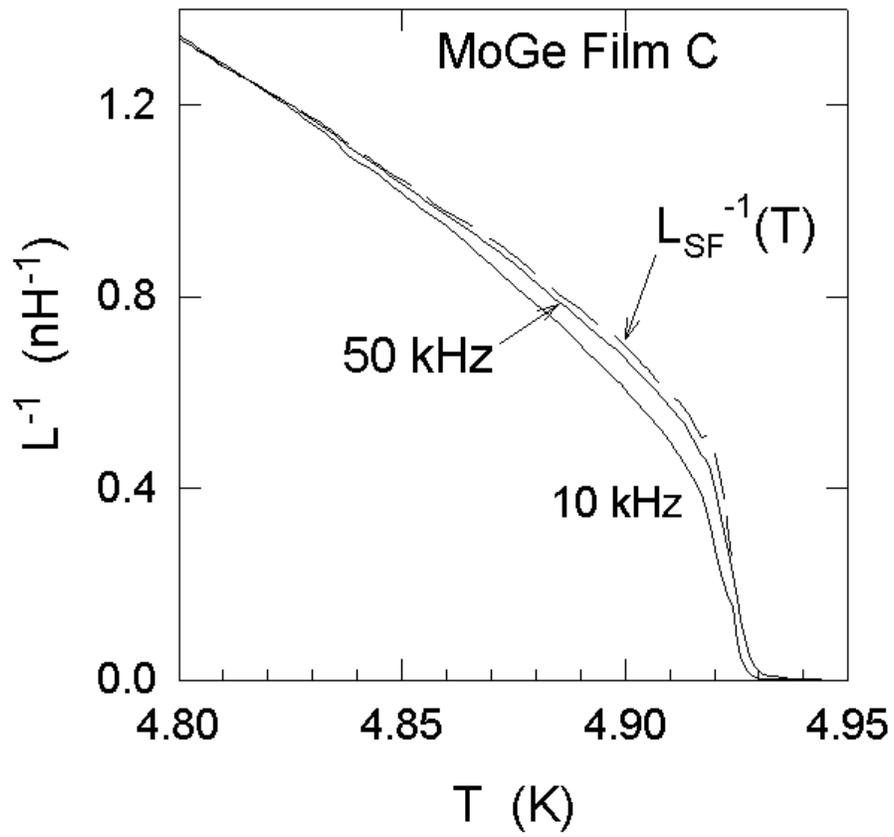

Figure 9.

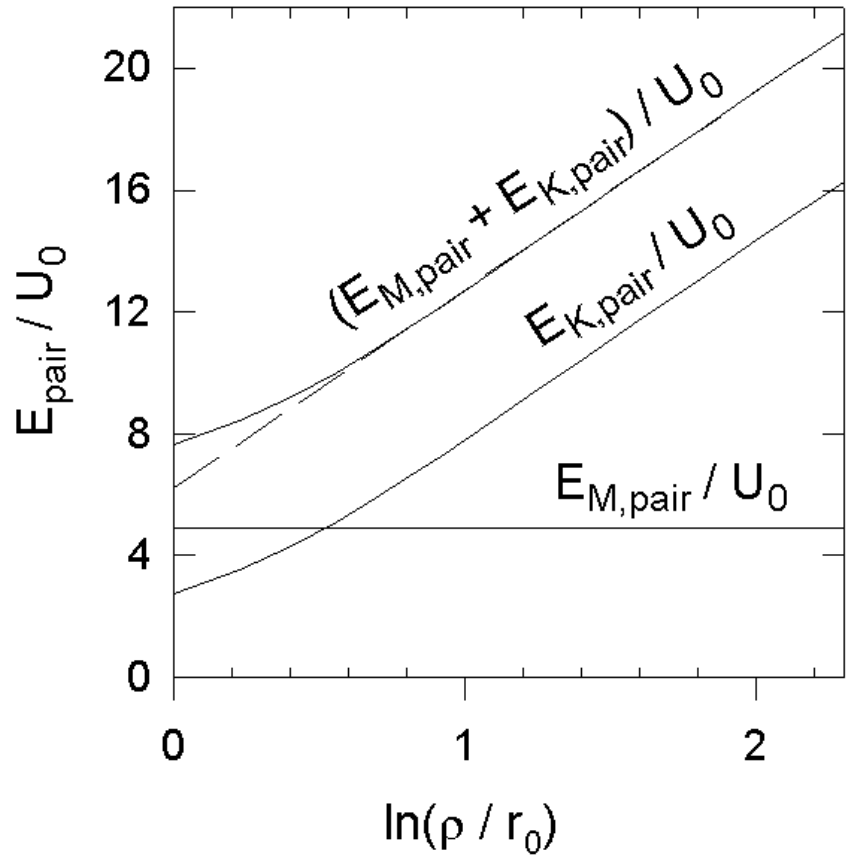

Figure 10.

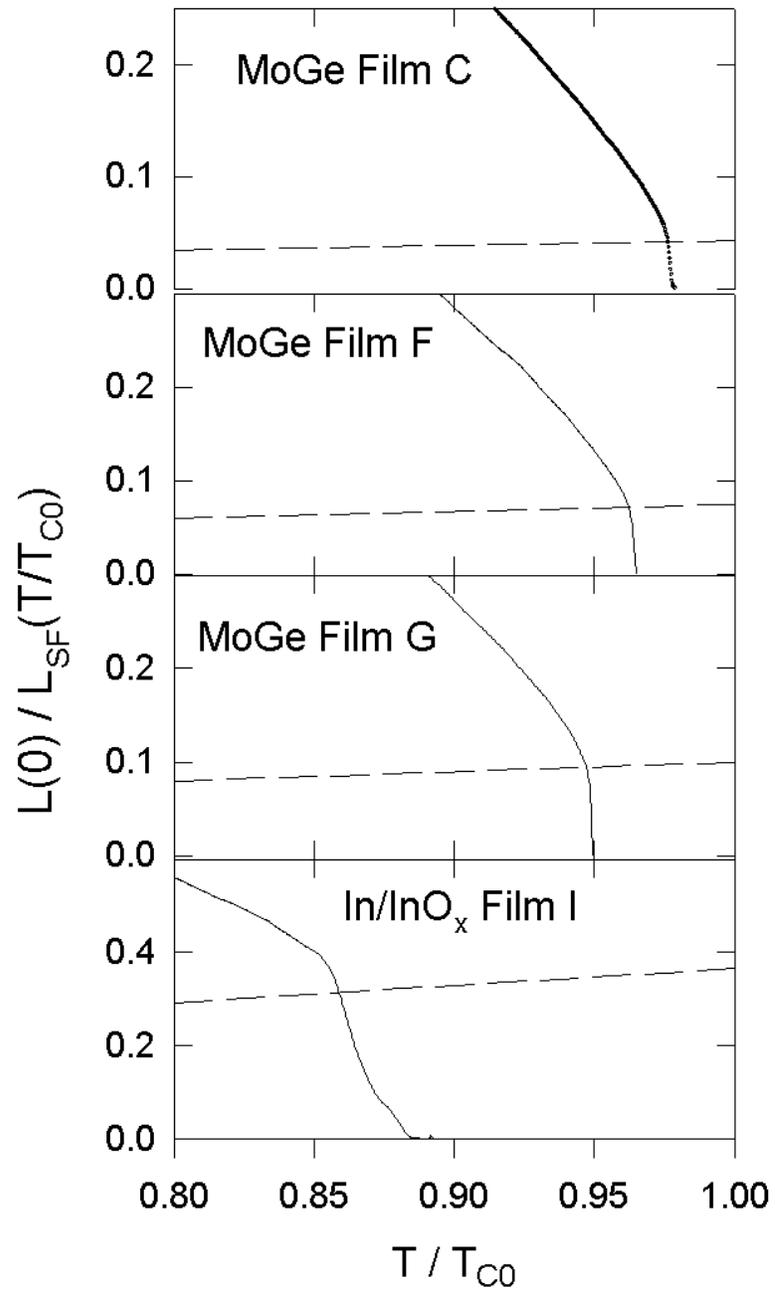